\documentstyle[titlepage,12pt]{article}

\title{Dynamics of Spreading of Small
Droplets of Chainlike Molecules on Surfaces}

\author{J. A. Nieminen$^1$ and T. Ala-Nissila$^{1,2}$ \\
\\ $^1$Department of Physics
\\ Tampere University
of Technology  \\ P.O.Box 692 \\ FIN - 33101 Tampere \\
Finland \\  \\
$^2$Research Institute for Theoretical
Physics \\ University of Helsinki \\
P.O. Box 9 (Siltavuorenpenger 20 C) \\
FIN - 00014 University of Helsinki \\
Finland \\  \\  \\ }

\date{January 3, 1994}

\begin{document}

\maketitle

\textheight 21cm
\textwidth 14.5cm

\oddsidemargin 0.96cm
\evensidemargin 0.96cm
\topmargin -0.31cm
\columnsep 0.5in
\raggedbottom

\baselineskip 24pt

\begin{abstract}

Dynamics of spreading of small droplets on
surfaces has been studied by the molecular dynamics
method. Simulations have been performed for
mixtures of solvent and dimer, and solvent and
tetramer droplets. For solvent particles and dimers, layering
occurs leading to stepped droplet shapes.
For tetramers such shapes occur
for relatively deep and strong surface potentials only.
For wider and
more shallow potentials, more rapid spreading and
rounded droplet shapes occur.
These results are in accordance with experimental
data on small non - volatile polymer droplets.


\end{abstract}

Experiments on the dynamics of spreading of tiny
non - volatile liquid droplets have
revealed a variety of
droplet shapes on the molecular scale
\cite{HesN,HesP,Alb,Val} which cannot be explained
by continuum hydrodynamic theories
\cite{Tan,deG,Joa,Leg}. These
experiments have created a need for
more microscopic models \cite{Leg}.
Some analytic models have been successful in describing
dynamical
layering \cite{Abr1,Abr2,deG90}. To obtain more
microscopic understanding,
molecular dynamics simulations have been
performed \cite{Hau,Ju}.
In particular, Nieminen
{\it et al.} \cite{Nie} simulated
Lennard - Jones (LJ) particles and observed the
appearance of a precursor film, dynamical layering,
and a crossover from almost linear ($\sim t^{0.9}$)
to $t^{0.5}$ behavior for the width of the
precursor \cite{Joa,Leg}.
\par

However, recent experiments using a variety
of polymer liquids
on energetically different surfaces show features which
are beyond the simple LJ model of a liquid
\cite{HesN,HesP,Alb,Val}.
For example, a droplet of squalane develops only
a tiny foot preceding the droplet \cite{HesP},
whereas in the case of polydimethylsiloxane (PDMS)
a monolayer thick precursor layer is seen \cite{HesP},
except on silver where the droplet has a spherical cap
and Gaussian tails at late times \cite{Alb}.
Furthermore, tetrakis (2 - ethylhexoxy) - silane develops
multiple (up to four)
precursor layers
and a tiny cap in the center of the droplet \cite{HesN}.
Ven\"al\"ainen {\it et al.} \cite{Ven}
have recently demonstrated that most droplet shapes
can be qualitatively
reproduced by an effective solid
on solid model where interparticle and substrate interactions
are varied.
However, there still remains an obvious need to understand
the role of microscopic solid - liquid and liquid - liquid
interactions, temperature, and,
in particular, the chainlike morphology
of the molecules. Although most experiments on tiny droplets
have been done with rather short molecules {\it beneath}
the static entanglement regime \cite{deG2}, recent work shows \cite{Alb}
how spherical droplet shapes can occur due to strong blocking effects
between short chainlike molecules.
This is the motivation for the present work, where we
present a study where the chainlike structure of the
spreading liquid {\it and} the influence
of the surface potential are taken into
account microscopically within the
molecular dynamics (MD) method.
\par

We consider a model system
of a mixture of ``solvent'' (single atom)
particles and ``polymers'' (short effective chainlike molecules) which
interact with each other via an LJ
potential, with potential
parameters $\sigma_f$ and $\epsilon_f$ for
the width and depth of
the potential, respectively. The substrate
is modeled by a flat
continuum LJ material \cite{Nie}.
For the substrate - fluid interaction,
the depth of the potential is
comparable to $\epsilon_s \sigma_s^3$ and the width to
$\sigma_s$ (see Ref. \cite{Nie}).
The $n$ - mers in the system
consist of $n$ LJ particles, which
are interconnected by a very rigid but
orientationally isotropic
harmonic oscillator pair potential
$ V_c = \frac{1}{2} k (r-r_0)^2, $
where $k = 100 \epsilon_f / \sigma_f^2$.
There is also an angle dependent potential
$ V_{\theta} = \epsilon_{\theta} (\cos\theta + 1), $
which favors linear molecules. Here we present results for
$\epsilon_{\theta} = 10 \epsilon_f $ which turns out to
be the most interesting case;
results for more flexible molecules
are similar, and will be published elsewhere \cite{Nie3}.
Due to the strong pair potential
between subsequent atoms comprising
an $n$ - mer, a strong repulsion
is created if any $n$ - mers try to spatially overlap.
\par

The dynamics of the system is described by
the usual equations
of motion \cite{Nie}:
$ dr_i / dt = p_i / m_i,$
$ dp_i / dt = \nabla _iV - \eta p_i, $ and
$ d\eta / dt = [\Sigma_i (p_i^2/m_i) - NkT]/NkT\tau^2, $
where $N$ is the number of degrees of freedom, $T$
the temperature for the
thermostat, and $\tau$ is a relaxation time.
The time scales are here
chosen in the same way as in Ref. \cite{Nie}.
Initially a ridge - shaped droplet is constructed
with periodic boundary conditions along
the direction of the ridge.
The spreading takes place in
the direction perpendicular to the ridge.
The equations of motion are
solved using modified
velocity Verlet algorithm
(see e.g. \cite{Ver} and \cite{Sut}).
\par

The initial configuration is constructed
so that the $n$ - mers are
bent 90 degrees at each joint, while
their directions are arbitrary.
To attain an optimum packing density
the system is compressed, after
which the system is allowed to
find a lower energy state.
After finding the best possible configuration
we switch on the
substrate interaction and allow the system to evolve.
\par

In the previous study of Nieminen {\it et al.}
\cite{Nie} the temperature
was $kT_s = 0.8 \epsilon_f$, which is
well above the triple
point of an LJ material \cite{Lad}.
In the present case the larger mass and
binding energy of the $n$ - mers
allows us to systematically
vary {\it both} the surface potential {\it and}
the temperature, and thus e.g. study
the effects of changing the
viscosity of the liquid.
In the present work, we employ
several sets of substrate
interaction parameters: $ \epsilon_s = 5 \epsilon_f$
and $\sigma_s = \sigma_f$
(used in Ref. \cite{Nie}), denoted as $S_I$;
$ \epsilon_s = \epsilon_f$ and $\sigma_s = 5.0\sigma_f$
($S_{II}$); $ \epsilon_s = 0.02\epsilon_f$
and $\sigma_s = 7.3\sigma_f$ ($S_{III}$); and
$ \epsilon_s = 0.002\epsilon_f$ and
$\sigma_s = 7.3\sigma_f$ ($S_{IV}$).
The differences between these
potentials can be characterized
by their steepness (gradient)
($ \sim \epsilon_s {\sigma_s}^2 $), depth, and width.
The potential $S_{II}$ is much steeper
and deeper than $S_I$, while
$S_{III}$ and $S_{IV}$ are both relatively
shallow, the latter being
particularly flat.
\par

First we describe results for mixtures of
solvent and dimer particles.
Qualitatively, a dimer concentration of
50\% or 100\% does not alter
the essential features of spreading as
compared to a pure solvent droplet.
Dimer droplets exhibit
well separated layers such as observed for
solvent droplets for
$S_{I}$, $S_{II}$ and $S_{III}$. However,
in order to clearly observe this,
a somewhat deeper substrate potential
$S_{II}$ must be used (for $S_I$ only
one precursor layer forms).
Despite this slight difference dimer
droplets show a rather
clear crossover from the almost linear
to $t^{0.5}$ spreading
as described in Ref. \cite{Nie} for solvent droplets.
The presence of dimers, however, makes
the spreading dynamics slightly slower.
\par

In the case of tetramers, effects arising
from the chain structure
of the liquid become more apparent.
The first striking feature
is a dynamical {\it ordering} of
the rather stiff chains along the ``flow''
of the spreading droplet, perpendicular
to the surface (Fig. 1).
The nature of layer separation is also
changed. Only for the very strong
surface potential $S_{II}$
a precursor layer with thickness of
about one monolayer is formed.
When surface potential becomes more shallow,
a distinct change in the droplet shapes occur.
For $S_{III}$ a vertically
continuous precursor layer whose
effective thickness is two layers is seen.
When the substrate potential
is further weakened to $S_{IV}$, the two layers tend to
merge together. This leads to a noticeably more rounded
droplet configurations than for dimers. The rate of spreading of the
droplet also becomes faster from $S_{II}$ to $S_{III}$.
This is consistent with the observation that
for a strongly attractive,
narrow surface potential the tetramers become
tightly packed {\it parallel} to the surface,
which slows down their
migration in the effectively
two dimensional layer \cite{Alb}.
Such layers also remain relatively
well separated.
On the other hand, a wider surface potential does not
force the tetramers in layers, thus allowing faster
dynamics and formation of a more rounded droplet configuration.
Mixing
of the layers in Fig. 1 is seen
which leads to a very complex structure within
the precursor layer.
When the substrate interaction is still weakened from
$(S_{III})$ to $(S_{IV})$, no further increase
in the rate of spreading is not seen.
\par

We have also studied
the effect of increasing
tetramer concentration starting from
a pure solvent droplet,
as shown in Fig. 2. This slows down
the spreading rate in analogy with the dimer case.
Most importantly,
we find that the width of the precursor film $w$
obeys a finite size scaling form as
a function of the number of
LJ particles $N$
\cite{Nie,Nie3} $w = t^x \ \phi (t/N^{y}),$
where $x \approx 9/10$ \cite{Joa}, and $y \approx 0.67 /x$.
This was checked for $2380  \le N \le 4080$.
This indicates that the spreading is
not purely diffusive but rather
caused by a combination of the
pressure of the cap and collective
migration of the tetramers
\cite{vii}. The scaling function $\phi(z) \sim const.$ for
$z \ll 1$, and $\phi(z) \sim z^{1/2 - x}$ for $z \gg 1$.
Details of these results will be published elsewhere \cite{Nie3}.
\par

Next we study the density profiles of the droplets in order
to make comparison with experiments \cite{HesN,HesP,Alb,Val}.
As seen in Fig. 3(a), the density profile
for a solvent droplet
shows a stepped precursor layer for $S_{II}$, indicating
separation of layers. Results for dimers are analogous, with slightly
more rounded step edges for weaker potentials.
For tetramers in the case of $S_{II}$,
the density profile shows a stepped shape, indicating
dynamical layering.
For weaker potentials the comparison of snapshots and
profiles shows mixing of the layers. For $S_{III}$
the precursor film on the whole is
fairly thick and the density continuosly increases
towards the centre. The density profile
is seen to develop towards a rounded cap \cite{Alb}, which
is most clearly seen in Fig. 3(b) for the case of $S_{IV}$.
Additional studies with more flexible chains and
octamers support these conclusions \cite{Nie3}.
\par

The profiles of Fig. 3 bear a remarkable
qualitative resemblence to some of the experimental ones.
The stepped profile of Fig. 3(a) is
qualitatively similar to that of a
tetrakis droplet consisting of
star - shaped molecules \cite{HesN}.
Our results indeed suggest that for spherical
or dimer molecules separation of layers
tends to occur.
On the other hand, the
profile of the tetramer droplet (Fig. 3(b)) resembles the
profiles of a PDMS droplet on silver \cite{Alb},
and on silicon
\cite{HesP}. These more rounded
shapes should occur for chainlike molecules,
when the surface
attraction is not strong. Furthermore,
on silicon a change
of the droplet from a rounded towards a stepped
shape was observed
on a ``high - energy'' surface. Our results suggest that this
indicates increased substrate attraction, which
tends to separate the layers.
A quantitative comparison of our results with
experiments is difficult, however, since
experimental droplets are microscopic in the vertical direction
{\it only}.
Nevertheless, our results support the conclusion
that {\it weakening}
the surface potential causes faster spreading and more rounded
droplet shapes for polymeric liquids.
Faster spreading of the
precursor film for weaker potentials
has also been predicted in Ref. \cite{Abr1}.
\par

Finally, we have also studied the effect of
temperature. As expected,
increasing the temperature increases the spreading rate.
For solvent droplets ($S_{II}$)
changing the temperature from
$T=T_S$ to $T=1.5T_S$ simply makes the steps more rounded.
For tetramer droplets in the case of $S_{III}$,
the effects are relatively small.
For $T=4T_S$ the profile shows
a minimum at the centre
of the droplet, after the vanishing of the cap.
This is followed by a fairly
flat rounded profile, as seen in Ref. \cite{Alb}.
\par

To summarize, in this work we have
studied effects arising from the chainlike molecular
structure within
spreading droplets,
and the form of the surface potential.
Our model simulations suggest that for relatively
strongly attracting
surfaces, dynamical layering and stepped
droplet shapes tend to occur.
On the other hand, the mixing of chainlike
molecules opposes
this trend leading to more rounded droplet
shapes on weaker
substrates. Our results also support
the conclusion that weaker
substrate attraction enhances the rate of spreading.
These results are in agreement with experiments
\cite{HesN,HesP,Alb,Val}.
We hope that our work inspires further
 systematic experiments
on the effects of chain lengths and surface
potentials on the
spreading of tiny droplets.

\par
Acknowledgements:
We wish to thank K. Kaski, R. Swendsen,
and O. Ven\"al\"ainen for useful discussions,
and S. Herminghaus
for a critical reading of the manuscript.
This work has been supported by the Academy of Finland.

\pagebreak

\Large
\noindent
{\bf Figure Captions} \\ \\

\normalsize

\noindent
Fig. 1(a)-(b). Snapshots of a projection of
a three dimensional
pure tetramer droplet for potential $S_{IV}$,
with $N = 2380$.
Times correspond to $50$ and $75$ in reduced
units \cite{Nie}.
Mixing of the layers is clearly visible.
Units in this and the other figures have been obtained from
$\sigma_f=2.6$ \AA. Due to the effective nature of the chainlike
molecules the length scales should be considered effective only.
\\

\noindent
Fig. 2. The effect of tetramer concentration $c$ on the
time evolution of the precursor width $w$
for potential $S_{II}$:
$c=0$ (full line), $c=0.5$ (dash-dotted line), and
$c=1$ (dotted line).
\\

\noindent
Fig. 3. Smoothed density profiles for
two different droplets. (a) A solvent
droplet for $S_{II}$ and (b) a tetramer
droplet for $S_{IV}$.
The relative times are $t_1$, $4t_1$, and $6t_1$.

\pagebreak

\end{document}